\documentclass[longauth]{aa}  

\usepackage{graphicx}
\usepackage{txfonts}
\usepackage{booktabs}
\usepackage{url}
\usepackage{xcolor}
\usepackage[breaklinks, colorlinks, citecolor=blue, linkcolor=blue]{hyperref}
\usepackage{scalerel}
\usepackage{tikz}
\usepackage{{multirow}}

\usepackage[english]{babel}
\usepackage[autostyle,english=british]{csquotes}

\newcommand{\emth}[1]{\ensuremath{#1}\xspace}

\newcommand{\tstar}{TOI-2266\xspace}
\newcommand{\tplanet}{TOI-2266 b\xspace}
\newcommand{\tic}{TIC~8348911\xspace}
\newcommand{\pytransit}{\textsc{PyTransit}\xspace}
\newcommand{\ldtk}{\textsc{LDTk}\xspace}
\newcommand{\spright}{\textsc{spright}\xspace}

\newcommand{\mps}{\emth{\mathrm{m\,s^{-1}}}}
\newcommand{\gcm}{\emth{\mathrm{g\,cm^{-3}}}}
\newcommand{\smass}{\emth{\mathrm{M_\star}}}
\newcommand{\srad}{\emth{\mathrm{R_\star}}}

\newcommand{\ktrue}{\ensuremath{k_\mathrm{true}}\xspace}

\newcommand{\teff}{\ensuremath{T_\mathrm{eff}}\xspace}

\newcommand{\rjup}{\ensuremath{R_\mathrm{Jup}}\xspace}
\newcommand{\msun}{\ensuremath{M_\odot}\xspace}
\newcommand{\rsun}{\ensuremath{R_\odot}\xspace}
\newcommand{\rearth}{\ensuremath{R_\oplus}\xspace}

\newcommand{\tess}{{TESS}\xspace}

\newcommand{\maroonx}{{MAROON-X}\xspace}
\newcommand{\kpf}{{KPF}\xspace}
\newcommand{\hipercam}{HiPERCAM\xspace}
\newcommand{\spirou}{{SPIRou}\xspace}

\newcommand{\operiod}{2.33\xspace}
\newcommand{\rmedian}{1.54\xspace}
\newcommand{\rerror}{0.09\xspace}

\newcommand{\orcidicon}[1]{
  \href{https://orcid.org/#1}{\includegraphics[scale=0.16]{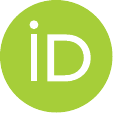}}
}

\usepackage[normalem]{ulem}

\begin{document} 

\title{\tplanet: a keystone super-Earth at the edge of the M~dwarf radius valley}
\titlerunning{\tplanet: a keystone super-Earth}

\author{H.~Parviainen\inst{\ref{iull},\ref{iiac}}\orcidicon{0000-0001-5519-1391}
    \and F.~Murgas\inst{\ref{iiac},\ref{iull}}\orcidicon{0000-0001-9087-1245}
    \and E.~Esparza-Borges\inst{\ref{iiac},\ref{iull}}\orcidicon{0000-0002-2341-3233}
    \and A.~Pel\'aez-Torres\inst{\ref{iiac},\ref{iull}}\orcidicon{0000-0001-9204-8498}
    \and E. Palle\inst{\ref{iiac},\ref{iull}}\orcidicon{0000-0003-0987-1593}
    \and R.~Luque\inst{\ref{uchicago}}\orcidicon{0000-0002-4671-2957}
    \and M.R.~Zapatero-Osorio\inst{\ref{csic}}\orcidicon{0000-0001-5664-2852}
    \and J.~Korth\inst{\ref{lund}}\orcidicon{0000-0002-0076-6239}
    \and A.~Fukui\inst{\ref{ikis},\ref{iiac}}\orcidicon{0000-0002-4909-5763}
    \and N.~Narita\inst{\ref{ikis},\ref{iabc},\ref{iiac}}\orcidicon{0000-0001-8511-2981}
    \and K.A.~Collins\inst{\ref{icfa}}\orcidicon{0000-0001-6588-9574} 
    \and V.J.S.~B\'ejar\inst{\ref{iiac},\ref{iull}}\orcidicon{0000-0002-5086-4232}
    \and G.~Morello\inst{\ref{chalmers},\ref{iiac}}\orcidicon{0000-0002-4262-5661}
    \and M.~Monelli\inst{\ref{iiac},\ref{iull}} 
    \and N.~Abreu~Garcia\inst{\ref{iiac},\ref{iull}}
    \and G.~Chen\inst{\ref{pmo}}
    \and N.~Crouzet\inst{\ref{ileiden}}\orcidicon{0000-0001-7866-8738}
    \and J.P.~de~Leon\inst{\ref{idmds}}\orcidicon{0000-0002-6424-3410}
    \and K.~Isogai\inst{\ref{iook},\ref{idmds}}\orcidicon{0000-0002-6480-3799}
    \and T.~Kagetani\inst{\ref{idmds}}\orcidicon{0000-0002-5331-6637}
    \and K.~Kawauchi\inst{\ref{iru}}\orcidicon{0000-0003-1205-5108}
    \and P.~Klagyivik\inst{\ref{iberlin}}
    \and T.~Kodama\inst{\ref{ikis}}\orcidicon{0000-0001-9032-5826}
    \and N.~Kusakabe \inst{\ref{iabc},\ref{inao}}\orcidicon{0000-0001-9194-1268}
    \and J.H.~Livingston\inst{\ref{iabc},\ref{inao},\ref{idas}}\orcidicon{0000-0002-4881-3620}
    \and P.~Meni\inst{\ref{iiac},\ref{iull}}\orcidicon{0009-0001-7943-0075}
    \and M.~Mori\inst{\ref{idmds}}\orcidicon{0000-0003-1368-6593}
    \and G.~Nowak\inst{\ref{inico},\ref{iiac},\ref{iull}} 
    \and M.~Tamura\inst{\ref{iutda},\ref{iabc},\ref{inao}}\orcidicon{0000-0002-6510-0681}
    \and Y.~Terada\inst{\ref{iaaas},\ref{intu}}\orcidicon{0000-0003-2887-6381}
    \and N.~Watanabe\inst{\ref{idmds}}\orcidicon{0000-0002-7522-8195}
    \and D.R.~Ciardi\inst{\ref{icaltech}}\orcidicon{0000-0002-5741-3047}
    \and M.B.~Lund\inst{\ref{icaltech}}\orcidicon{0000-0003-2527-1598}
    \and J.L.~Christiansen\inst{\ref{icaltech}}\orcidicon{0000-0002-8035-4778}
    \and C.D.~Dressing\inst{\ref{iberkeley}}\orcidicon{0000-0001-8189-0233}
    \and S.~Giacalone\inst{\ref{iberkeley}}\orcidicon{0000-0002-8965-3969}
    \and A.B.~Savel\inst{\ref{imary}}\orcidicon{0000-0002-0786-7307}
    \and L.~Hirsch\inst{\ref{istanford}}\orcidicon{0000-0001-8058-7443}
    \and S.G.~Parsons\inst{\ref{isheffield}}
    \and P.~Brown\inst{\ref{ivandy}}\orcidicon{0000-0002-3481-9052} 
    \and K.I.~Collins\inst{\ref{igmu}}\orcidicon{0000-0003-2781-3207} 
    \and K.~Barkaoui\inst{\ref{astro_liege},\ref{MIT},\ref{iull}}\orcidicon{0000-0003-1464-9276}
    \and M.~Timmermans\inst{\ref{astro_liege}}
    \and M.~Ghachoui\inst{\ref{ioukai},\ref{astro_liege}}\orcidicon{0000-0003-3986-0297}
    \and A.~Soubkiou\inst{\ref{ioukai},\ref{iportoa},\ref{iportob}}
    \and Z.~Benkhaldoun\inst{\ref{ioukai}}\orcidicon{0000-0001-6285-9847}
    %
    \and S.~McDermott\inst{\ref{iproto}}
    \and T.~Pritchard\inst{\ref{igoddard}}
    \and P.~Rowden\inst{\ref{iras}}
    %
    \and S.~Striegel\inst{\ref{iames}}
    \and T.~Gan\inst{\ref{itca}}\orcidicon{0000-0002-4503-9705}
    \and{K.~Horne}\inst{\ref{supa}} 
    \and E.L.N.~Jensen\inst{\ref{isc}}\orcidicon{0000-0002-4625-7333} 
    \and R.P.~Schwarz\inst{\ref{icfa}}\orcidicon{0000-0001-8227-1020} 
    \and{A.~Shporer}\inst{\ref{imit}}\orcidicon{0000-0002-1836-3120} 
    \and G.~Srdoc\inst{\ref{koto}} 
    %
    \and S.~Seager\inst{\ref{imit},\ref{imitate},\ref{imiteap}}\orcidicon{0000-0002-6892-6948}
    \and J.N.~Winn\inst{\ref{iprinceton}}\orcidicon{0000-0002-4265-047X}
    \and J.M.~Jenkins\inst{\ref{iames}}\orcidicon{0000-0002-4715-9460}
    \and G.~Ricker\inst{\ref{imit}}\orcidicon{0000-0003-2058-6662}
    \and R.~Vanderspek\inst{\ref{imit}}\orcidicon{0000-0001-6763-6562}
    %
    \and D.~Dragomir\inst{\ref{iunm}}
    %
}

\institute{
    Dept. Astrof\'isica, Universidad de La Laguna (ULL), E-38206 La Laguna, Tenerife, Spain\label{iull}
    \and Instituto de Astrof\'isica de Canarias (IAC), E-38200 La Laguna, Tenerife, Spain\label{iiac}
    \and Department of Astronomy \& Astrophysics, University of Chicago, Chicago, IL 60637, USA\label{uchicago}
    \and Centro de Astrobiologia (CSIC-INTA), Carretera de Ajalvir km 4, 28850 Torrejon de Ardoz, Madrid, Spain\label{csic}
    \and Lund Observatory, Division of Astrophysics, Department of Physics, Lund University, Box 43, 22100 Lund, Sweden \label{lund}
    \and Komaba Institute for Science, The University of Tokyo, 3-8-1 Komaba, Meguro, Tokyo 153-8902, Japan \label{ikis}
    \and Astrobiology Center, 2-21-1 Osawa, Mitaka, Tokyo 181-8588, Japan \label{iabc}
    \and Center for Astrophysics ${\rm \mid}$ Harvard {\rm \&} Smithsonian, 60 Garden Street, Cambridge, MA 02138, USA \label{icfa}
    \and Department of Space, Earth and Environment, Chalmers University of Technology, SE-412 96 Gothenburg, Sweden\label{chalmers}
    \and Key Laboratory of Planetary Sciences, Purple Mountain Observatory, Chinese Academy of Sciences, Nanjing 210023, China \label{pmo}
    \and Leiden Observatory, Leiden University, P.O. Box 9513, 2300 RA Leiden, The Netherlands \label{ileiden}
    \and Department of Multi-Disciplinary Sciences, Graduate School of Arts and Sciences, The University of Tokyo, 3-8-1 Komaba, Meguro, Tokyo 153-8902, Japan \label{idmds} 
    \and Okayama Observatory, Kyoto University, 3037-5 Honjo, Kamogatacho, Asakuchi, Okayama 719-0232, Japan \label{iook}
    \and Department of Physical Sciences, Ritsumeikan University, Kusatsu, Shiga 525-8577, Japan \label{iru}
    \and Freie Universit\"at Berlin, Institute of Geological Sciences, Malteserstr. 74-100, 12249 Berlin, Germany \label{iberlin}
    \and National Astronomical Observatory of Japan, 2-21-1 Osawa, Mitaka, Tokyo 181-8588, Japan \label{inao}
    \and Department of Astronomical Science, The Graduated University of Advanced Studies, SOKENDAI, 2-21-1, Osawa, Mitaka, Tokyo, 181-8588, Japan \label{idas}
    \and Institute of Astronomy, Faculty of Physics, Astronomy and Informatics, Nicolaus Copernicus University, Grudzi\c{a}dzka 5, 87-100 Toru\'n, Poland \label{inico}
    \and Department of Astronomy, The University of Tokyo, 7-3-1 Hongo, Bunkyo-ku, Tokyo 113-0033, Japan \label{iutda}
    \and Institute of Astronomy and Astrophysics, Academia Sinica, P.O. Box 23-141, Taipei 10617, Taiwan, R.O.C. \label{iaaas}
    \and Department of Astrophysics, National Taiwan University, Taipei 10617, Taiwan, R.O.C. \label{intu}
    \and NASA Exoplanet Science Institute - Caltech/IPAC, Pasadena, CA 91125, USA \label{icaltech}
    \and Department of Astronomy, University of California Berkeley, Berkeley, CA 94720, USA \label{iberkeley}
    \and Department of Astronomy, University of Maryland, College Park, MD 20742, USA \label{imary}
    \and Kavli Center for Particle Astrophysics and Cosmology, Stanford University, Stanford, CA 94305, USA) \label{istanford}
    \and Department of Physics and Astronomy, University of Sheffield, Sheffield, S3 7RH, UK \label{isheffield}
    \and Department of Physics and Astronomy, Vanderbilt University, Nashville, TN 37235, USA \label{ivandy}
    \and George Mason University, 4400 University Drive, Fairfax, VA, 22030 USA \label{igmu}
    \and Astrobiology Research Unit, Universit\'e de Li\`ege, All\'ee du 6 Ao\^ut 19C, B-4000 Li\`ege, Belgium \label{astro_liege}
    \and Department of Earth, Atmospheric and Planetary Science, Massachusetts Institute of Technology, 77 Massachusetts Avenue, Cambridge, MA 02139, USA \label{MIT}
    \and Oukaimeden Observatory, High Energy Physics and Astrophysics Laboratory, Faculty of sciences Semlalia, Cadi Ayyad University, Marrakech, Morocco \label{ioukai}
    \and Departamento de Fisica e Astronomia, Faculdade de Ciencias, Universidade do Porto, Rua do Campo Alegre, 4169-007 porto, Portugal \label{iportoa}
    \and Instituto de Astrofisica e Ciencias do Espaco, Universidade do Porto, CAUP, Rua das Estrelas, 150-762 Porto, Portugal \label{iportob}
    \and Proto-Logic LLC, 1718 Euclid Street NW, Washington, DC 20009, USA \label{iproto}
    \and NASA Goddard Space Flight Center, 8800 Greenbelt Road, Greenbelt, MD 20771, USA \label{igoddard}
    \and Royal Astronomical Society, Burlington House, Piccadilly, London W1J 0BQ, United Kingdom \label{iras}
    \and SETI Institute, Mountain View, CA 94043 USA/NASA Ames Research Center, Moffett Field, CA 94035 USA \label{iames}
    \and Department of Astronomy and Tsinghua Centre for Astrophysics, Tsinghua University, Beijing 100084, China \label{itca}
    \and SUPA Physics and Astronomy, University of St. Andrews, Fife, KY16 9SS Scotland, UK \label{supa}
    \and Dept.\ of Physics \& Astronomy, Swarthmore College, Swarthmore PA 19081, USA \label{isc}
    \and Patashnick Voorheesville Observatory, Voorheesville, NY 12186, USA \label{pvo}
    \and Department of Physics and Kavli Institute for Astrophysics and Space Research, Massachusetts Institute of Technology, Cambridge, MA 02139, USA \label{imit}
    \and Kotizarovci Observatory, Sarsoni 90, 51216 Viskovo, Croatia \label{koto}
    \and Department of Aeronautics and Astronautics, Massachusetts Institute of Technology, Cambridge, MA 02139, USA \label{imitate}
    \and Department of Earth, Atmospheric, and Planetary Sciences, Massachusetts Institute of Technology, Cambridge, MA 02139, USA \label{imiteap}
    \and Department of Astrophysical Sciences, Princeton University, Princeton, NJ 08544, USA \label{iprinceton}
    \and Department of Physics and Astronomy, University of New Mexico, 1919 Lomas Blvd NE, Albuquerque, NM 87131, USA \label{iunm}
}

\date{Received MM DD, 20XX; accepted MM DD, 20XX}

\abstract{We validate the Transiting Exoplanet Survey Satellite (\tess) object of interest TOI-2266.01 (\tic) as a small
transiting planet (most likely a super-Earth) orbiting a faint M5 dwarf ($V=16.54$) on a \operiod~d orbit. The validation is based on an approach where multicolour transit light curves are used to robustly estimate the upper limit of the transiting object's radius. Our analysis uses SPOC-pipeline \tess light curves from Sectors 24, 25, 51, and 52, simultaneous multicolour transit photometry observed with MuSCAT2, MuSCAT3, and \hipercam, and additional transit photometry observed with the LCOGT telescopes. \tplanet is found to be a planet with a radius of $\rmedian\pm\rerror$~\rearth, which locates it at the edge of the transition zone between rocky planets, water-rich planets, and sub-Neptunes (the so-called M~dwarf radius valley). The planet is amenable to ground-based radial velocity mass measurement with red-sensitive spectrographs installed in large telescopes, such as \maroonx and Keck Planet Finder (\kpf), which makes it a valuable addition to a relatively small population of planets that can be used to probe the physics of the transition zone. Further, the planet's orbital period of \operiod days places it inside a \enquote{keystone planet} wedge in the period-radius plane where competing planet formation scenarios make conflicting predictions on how the radius valley depends on the orbital period. This makes the planet also a welcome addition to the small population of planets that can be used to test small-planet formation scenarios around M~dwarfs.}

\maketitle

\section{Introduction} 
\label{sec:introduction}

The radius distribution for Earth-to-Neptune-sized exoplanets on short-period orbits around M~dwarfs is bimodal, seeming to imply the existence of two planet populations with distinct physical properties \citep{Cherubim2022, Luque2022a, Luque2021, Cloutier2021, VanEylen2021, Cloutier2020b, Cloutier2020a, Cloutier2020}. This bimodality is similar to what has been observed for planets orbiting FGK-stars \citep{Mayo2018, Fulton2018, Fulton2017}, but the minimum between the two modes, also known as the radius valley, is located at $1.4-1.7$~\rearth for M dwarfs, a somewhat smaller radius than observed for the FGK-star radius valley ($1.7-2.0$~\rearth). 

The smaller-radius population of planets is expected to consist of rocky planets with negligible atmospheres (sub-Earths, Earths, and super-Earths), while the larger-radius population has been considered to consist of Neptune-like ice giants with extended H/He envelopes (sub-Neptunes). Water-rich planets (water worlds) with a water-to-rock ratio close to unity but lacking a significant H/He envelope have also been suggested as a third major population between rocky planets and sub-Neptunes \citep{Zeng2019}, but observational evidence supporting this was limited until recent work by \citet{Luque2022}. 

In their study, \citeauthor{Luque2022} focused on a set of small planets orbiting M~dwarfs on periods shorter than 35 days with masses and radii estimated to a precision of 25\% and 8\% or better, respectively. They found that the density distribution for small planets features three modes agreeing with the densities predicted for rocky planets, water worlds, and sub-Neptunes by \citet{Zeng2019}. From this perspective, the bimodal radius distribution would correspond to a projection of the density distribution blending the water worlds and sub-Neptunes together.

The result by \citet{Luque2022} is based on a small number of well-characterised planets, and the statistical significance of the hypothesis of three main planet-type populations can be improved by discovering more Earth-to-Neptune-sized planets around M~dwarfs. The Transiting Exoplanet Survey Satellite \citep[\tess;][]{Ricker2014} has identified several hundred planet candidates fitting the period and radius criteria used by \citet{Luque2022},\!\footnote{The TESS Project Candidate table in the NASA Exoplanet Archive contains 172 open candidates (disposition PC) around M~dwarfs with periods shorter than 35~d and radii smaller than 5~R$_\oplus$ at the time of writing (2022~October~21).} but some of these candidates are false positives, and instruments capable of carrying out radial velocity (RV) mass estimation of small planets around M~dwarfs are few and in high demand. Consequently, the first step in the process of identifying the main small-planet population types is to validate and characterise planet candidates amenable to RV mass estimation.

The planets used to study the small-planet populations can also be used to probe how small planets form around M~dwarfs \citep{Burn2021, Stefansson2020, Lopez2018}. The formation of short-period non-rocky planets (water worlds and sub-Neptunes) is roughly understood since these planets are expected to have formed originally beyond the protoplanetary disk ice line to accrete the water and gases that make them what they are, after which they have migrated inwards to their current orbits. However, the formation of rocky planets is still an open question with two proposed main competing formation pathways: gas-depleted formation or formation through thermally-driven mass loss. The gas-depleted formation proposes that rocky planets and water- and gas-rich planets consist of two separate planet populations that formed at different times. In contrast, the thermally-driven mass loss scenario proposes that rocky planets are basically sub-Neptune cores stripped of their H/He envelopes. These formation scenarios (discussed in more detail later in Sect.~\ref{sec:discussion}) lead to conflicting predictions on how the upper limit of the rocky planet size (that is, the centre of the radius valley) depends on the orbital period of the planet. Planets that are located in the area in the period-radius plane where the predictions from the two formation scenarios disagree (named \enquote{keystone planets} by \citealt{Cloutier2021}) can be used to probe which of the pathways is the dominant one.  

Here we report the validation and characterisation of \tplanet, a small transiting planet ($\rmedian \pm \rerror\,\rearth$), orbiting a faint M5 dwarf (\tic, see Table~\ref{tbl:star}) on a \operiod~d orbit.
The \emph{Science Processing Operations Center} (SPOC) located at NASA Ames Research Center conducted a transit search of Sector 24 on 2020~August~2 with an adaptive, noise-compensating matched filter \citep{Jenkins2002, Jenkins2010, Jenkins2016} producing a Threshold Crossing Event (TCE) for which an initial limb-darkened transit model was fitted \citep{Li2019} and a suite of diagnostic tests were conducted to help make or break the planetary nature of the signal \citep{Twicken2018}. The \tess Science Office (TSO) reviewed the vetting information and issued an alert on 30 September 2020 \citep{Guerrero2021}. The transit signature passed all the diagnostic tests presented in the Data Validation reports. The host star is located within $1.95\pm3.90$~arcsec of the source of the transit signal. The planet candidate was later followed up from the ground using multicolour transit photometry and low-resolution spectroscopy. The validation is carried out using the multicolour transit validation approach described in \citet{Parviainen2019} and applied later in \citet{Parviainen2020}, \citet{Parviainen2021}, \citet{Esparza-Borges2022}, and \citet{Morello2023}. The analyses and data discussed in this paper are publicly available from GitHub.\!\footnote{\url{https://github.com/hpparvi/parviainen_2021_toi_2266}}

\begin{table}[t]    
    \caption{\tstar identifiers, coordinates, properties, and magnitudes.}
    \centering
    \begin{tabular*}{\columnwidth}{@{\extracolsep{\fill}} llrr}
        \toprule\toprule
        \multicolumn{4}{l}{\emph{Main identifiers}}     \\
        \midrule     
        TIC   & & \multicolumn{2}{r}{8348911} \\
        2MASS & & \multicolumn{2}{r}{J16210714+3134367}   \\ 
        Gaia DR2 & & \multicolumn{2}{r}{1319243773843954304} \\
        \\
        \multicolumn{4}{l}{\emph{Equatorial coordinates}}     \\
        \midrule            
        RA \,(J2000) &  & \multicolumn{2}{r}{$16^h\,21^m\,07\fs21$}            \\
        Dec (J2000)  &  & \multicolumn{2}{r}{$31\degr\,34\arcmin\,37\farcs35$}  \\
        \\     
        \multicolumn{4}{l}{\emph{Stellar parameters }} \\
        \midrule
        Eff. temperature & \teff & [K] & $3240 \pm 160$\\
        Mass    & \smass &[\msun]  & $0.23 \pm 0.02$\\  
        Radius  & \srad  &[\rsun]  & $0.24 \pm 0.01$\\ 
        Parallax & & [mas] & $19.29 \pm 0.02$\\
        Distance & & [pc] & $51.72 \pm 0.06$\\
        Age & & [Myr] & $>300$ \\
        Spectral type & & & M5.0$^{+0.5}_{-0.5}$ \\
        \\
        \multicolumn{4}{l}{\emph{Magnitudes}} \\
        \midrule              
        \centering
        
        Filter & & Magnitude       & Uncertainty  \\
        \midrule     
        TESS & & 13.5042 & 0.0076 \\
        B  & & 17.962 & 0.162 \\
        V  & & 16.54 & 0.200 \\
        Gaia & & 14.8319 & 0.0005 \\
        u  & & 19.745 & 0.031 \\
        g  & & 17.018 & 0.004 \\ 
        r  & & 15.544 & 0.004 \\
        i  & & 14.072 & 0.004 \\
        z  & & 13.291 & 0.004 \\
        J  & & 11.844 & 0.022 \\
        H  & & 11.283 & 0.023 \\
        K  & & 11.017 & 0.021 \\
        \bottomrule
    \end{tabular*}
    \tablefoot{The stellar properties except the distance are based on a spectrum observed with ALFOSC, and their derivation is described in Sect.~\ref{sec:observations.alfosc_details}; the distance is from the tabulations by \citet{Bailer-Jones2021}; and the magnitudes are from ExoFOP.}
    \label{tbl:star}  
\end{table}

\section{Stellar characterisation} 
\label{sec:observations.alfosc_details} 
We obtained an optical low-resolution spectrum of \tstar{} with the Alhambra Faint Object Spectrograph and
Camera (ALFOSC) mounted at the 2.56~m Nordic Optical Telescope (NOT) on the Roque de los Muchachos
Observatory (ORM) on 2021~July~1~UT. ALFOSC is equipped with a 2048$\times$2064 CCD detector with a pixel scale of 0.2138 \arcsec
pixel$^{-1}$. We used grism number 5 and an horizontal long slit with a width of 1.0\arcsec\!, which yield a nominal
spectral dispersion of 3.53 \AA~pixel$^{-1}$ and a usable wavelength space coverage between 5000 and 9400~\AA. Two
spectra of 900~s each were acquired at parallactic angle and airmass of 1.03. We also observed a spectrophotometric
standard star {BD+17 4708} with the same instrumental setup as \tstar{}, with an exposure time of 15~s, and at an
airmass of 1.02. Raw images were reduced following standard procedures at optical wavelengths: bias subtraction,
flat-fielding using dome flats, and optimal extraction using appropriate packages within the IRAF\footnote{Image
Reduction and Analysis Facility (IRAF) is distributed by the National Optical Astronomy Observatories, which are
operated by the Association of Universities for Research in Astronomy, Inc., under contract with the National Science
Foundation.}  environment. Wavelength calibration was performed with a precision of 0.65 \AA~using He\,{\sc i} and
Ne\,{\sc i} arc lines observed on the same night. The instrumental response was corrected using observations of the
standard star. Because the primary target and the standard star were observed close in time and at a similar airmass,
we corrected for telluric lines absorption by dividing the target data by the spectrum of the standard normalised to
the continuum. 

The estimation of the stellar parameters (spectral type, effective temperature, and stellar mass and radius) was carried out as in \citet{Parviainen2021, Parviainen2020b} based on tabulations by \citet{Schweitzer2019} and \citet{Mann2019}, and the parameters are listed in Table~\ref{tbl:star}. We used the reference spectra of \citet{Kesseli2017} for the spectral classification, and the spectrum is compatible with solar metallicity.
Further, the astrometry of \tstar is incompatible with membership in young stellar moving groups independently of its radial velocity, so the star is likely not young (age > 300 Myr). This is also evident from the strength of the atomic lines (particularly K I and Na I) from the ALFOSC spectrum.

\section{High-resolution imaging}
\subsection{Palomar observations} 

\begin{figure*}
    \centering
    \includegraphics[width=\textwidth]{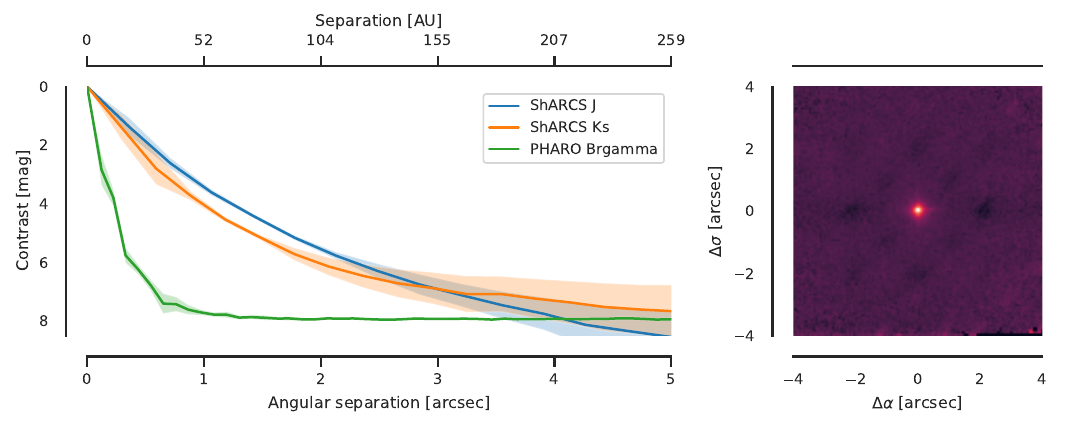}
    \caption{Contrast with its 1-$\sigma$ uncertainties as a function of separation and angular separation in Brgamma, 
    J and Ks (left) and a high-resolution PHARO image (right).}
    \label{fig:high_res_imaging}
\end{figure*}

As part of our standard process for validating transiting exoplanets to assess the possible contamination of bound or
unbound companions on the derived planetary radii \citep{ciardi2015}, we observed \tstar with high-resolution
near-infrared adaptive optics (AO) imaging at Lick and Palomar Observatories.  While the Palomar observations provided
higher resolution and sensitivity, the Lick observations provided multiple filters.  Neither set of observations detected
additional stars; additionally, the Gaia DR3 astrometry is consistent with the star being single with an astrometric
excess noise value of $<1.4$ ($RUWE = 1.03$).

The Palomar Observatory observations were made with the PHARO instrument
\citep{hayward2001} behind the natural guide star AO system P3K \citep{dekany2013} on 2021~February~24~UT in a standard
5-point quincunx dither pattern with steps of 5\arcsec\ in the narrow-band $Br-\gamma$ filter $(\lambda_o = 2.1686;
\Delta\lambda = 0.0326~\mu$m).  Each dither position was observed three times, offset in position from each other by
0.5\arcsec\ for a total of 15 frames; with an integration time of 10 seconds per frame, the total on-source time was 150
seconds. PHARO has a pixel scale of $0.025\arcsec$ per pixel for a total field of view of $\sim25\arcsec$.

The AO data were processed and analysed with a custom set of IDL tools.  The science frames were flat-fielded and
sky-subtracted. The flat fields were generated from a median average of dark subtracted flats taken on the sky. The flats
were normalised such that the median value of the flats was unity. The sky frames were generated from the median average
of the 15 dithered science frames; each science image was then sky-subtracted and flat-fielded. The reduced science
frames were combined into a single combined image using an intra-pixel interpolation that conserves flux, shifts the
individual dithered frames by the appropriate fractional pixels, and median-adds the frames. The final resolution of
the combined dithers was determined from the full-width half-maximum of the point spread function: 0.108\arcsec.  The
sensitivities of the final combined AO image were determined by injecting simulated sources azimuthally around the
primary target every $20^\circ $ at separations of integer multiples of the central source's FWHM \citep{furlan2017}. The
brightness of each injected source was scaled until standard aperture photometry detected it with $5\sigma $
significance. The resulting brightness of the injected sources relative to \tstar set the contrast limits at that
injection location. The final $5\sigma $ limit at each separation was determined from the average of all of the
determined limits at that separation, and the uncertainty on the limit was set by the rms dispersion of the azimuthal
slices at a given radial distance (Fig.~\ref{fig:high_res_imaging}).

\subsection{Lick observations} 
We observed TIC 8348911 on 2021~March~28~UT using the ShARCS camera on the Shane 3-meter
telescope at Lick Observatory \citep{2012SPIE.8447E..3GK, 2014SPIE.9148E..05G, 2014SPIE.9148E..3AM}. The observation was
taken with the Shane adaptive optics system in natural guide star mode. The final images were constructed using sequences
of images taken in a 4-point dither pattern with a separation of 4$\arcsec$ between each dither position. Two image
sequences were taken of this star: one with a $Ks$ filter ($\lambda_0 = 2.150$ $\mu$m, $\Delta \lambda = 0.320$ $\mu$m)
and one with a $J$ filter ($\lambda_0 = 1.238$ $\mu$m, $\Delta \lambda = 0.271$ $\mu$m), both of which used an exposure
time of 60 s at each dither position. A more detailed description of the observing strategy and reduction procedure can
be found in \cite{2020AJ....160..287S}. The contrast curves extracted from these observations are shown in
Fig.~\ref{fig:high_res_imaging}. We find no nearby stellar companions within our detection limits.

\section{Transit light curve analysis}
\label{sec:transits}
\subsection{Observations}
\label{sec:transits.observations}

\subsubsection{\tess photometry}
\label{sec:transits.observations.tess}


\tess observed 31 full transits of \tplanet during Sectors 24, 25, 51, and 52 with a two-minute cadence. We chose to use the Presearch Data Conditioning (PDC) light curves \citep{Stumpe2014, Stumpe2012, Smith2012a} produced by the SPOC pipeline, but, as in \citet{Parviainen2020} and \citet{Parviainen2021}, we add back the crowding correction (\enquote{CROWDSAP}) removed by the SPOC pipeline since the crowding correction could introduce a bias into our parameter estimation if the crowding were to be overestimated by the SPOC pipeline.
The final \tess photometry used in the transit analysis consists of 31 7.2~hour-long windows centred around each transit based on the linear ephemeris, and each window was normalised to its median out-of-transit level assuming a transit duration of 2.4~h. The photometry has an average point-to-point (ptp) scatter of 11.5~parts per thousand.

\subsubsection{MuSCAT2 photometry}
\label{sec:transits.observations.muscat2}

We observed five full transits of \tplanet simultaneously in $g$, $r$, $i$, and $z_\mathrm{s}$ bands with the MuSCAT2 multicolour imager \citep{Narita2018} installed at the 1.52~m Telescopio Carlos Sanchez (TCS) in the Teide Observatory, Spain, on the nights of 2021 February 9, 2021 March 3, 2021 June 24, 2021 July 8, and 2021 July 15. The exposure times were optimised for each night and CCD and varied from 30 to 120 seconds. The observing conditions were mostly good through all the nights, but we decided to discard the $g$ band photometry because of the low signal-to-noise ratio, and we also discarded the $r$ band photometry observed on a night with anomalously bad seeing.
The photometry was carried out using standard aperture photometry calibration and reduction steps with a dedicated MuSCAT2 photometry pipeline, as described in \citet{Parviainen2020}. Values for x- and y- centroid shifts, airmass, and PSF width were also extracted and stored to be used in the analysis as baseline model components.

\subsubsection{LCOGT 1\,m and TRAPPIST photometry}
\label{sec:transits.observations.lco}

We observed three transits of \tplanet in Sloan $i'$ band from the Las Cumbres Observatory Global Telescope \citep[LCOGT;][]{Brown2013} 1.0\,m network. A full transit was observed on 2021~March~5 from the McDonald Observatory node, and near-full and full transits were observed on 2021~March~26 and 2021~April~16, respectively, from the Cerro Tololo Inter-American Observatory node. We used the {\tt TESS Transit Finder}, which is a customised version of the {\tt Tapir} software package \citep{Jensen2013}, to schedule our transit observations. The $4096\times4096$ LCOGT SINISTRO cameras have an image scale of $0\farcs389$ per pixel, resulting in a $26\arcmin\times26\arcmin$ field of view. The images were calibrated by the standard LCOGT {\tt BANZAI} pipeline \citep{McCully2018}, and photometric data were extracted with {\tt AstroImageJ} \citep{Collins2017}. The images were focused and have typical stellar point-spread-functions with a full-width-half-maximum (FWHM) of roughly $2 \arcsec$, and circular apertures with radius $4\arcsec$ were used to extract the differential photometry.

Several transits of \tplanet were also observed with the TRAPPIST-South and TRAPPIST-North telescopes. However, the SNRs from these observations were too low to include in the analysis.

\subsubsection{LCOGT MuSCAT3 photometry} \label{sec:observations.lcoM3} A full transit of \tplanet was observed simultaneously in Sloan $g$, $r$, $i$, and Pan-STARRS $z$-short bands on 2021~May~23 using the LCOGT 2\,m Faulkes Telescope North at Haleakala Observatory on Maui, Hawai'i. The telescope is equipped with the MuSCAT3 multi-band imager \citep{Narita2020}. The images were calibrated using the standard LCOGT BANZAI pipeline, and photometric data were extracted using {\tt AstroImageJ}. The images were mildly defocused and had typical stellar point spread functions (PSFs) with FWHM of $\sim 2\farcs5$, and circular apertures with radius $4\arcsec$ were used to extract the differential photometry.

\subsubsection{\hipercam photometry} A full transit of \tplanet was observed simultaneously in  $u$, $g$, $r$, $i$, and $z$ with the High PERformance CAMera (\hipercam, \citealp{Dhillon2021}) mounted on the 10.4 m Gran Telescopio Canarias (GTC) on ORM on 2021~August~5. \hipercam is a multicolour imager composed of 5 CCD cameras capable of obtaining simultaneous observations in $u$, $g$, $r$, $i$, and $z$. Each camera has a field of view of $2.8\arcmin \times 1.4\arcmin$ with a pixel scale of $0.081$\arcsec{} pixel$^{-1}$. The exposure time was set to 1.69 seconds for all bands, and the data acquisition started at $\sim$22:20 UT (airmass 1.10) and ended at $\sim$01:00 UT (airmass 1.88).

\subsection{Multicolour planet candidate validation and system characterisation} 
\label{sec:transits.analysis}

We modelled the \tess light	curves simultaneously with the MuSCAT2, \hipercam, and LCOGT light curves following the approach described in \citet{Parviainen2019} and used in \citet{Parviainen2020}, \citet{Parviainen2021}, \citet{Esparza-Borges2022}, and \citet{Morello2023}.  Briefly, multicolour planet candidate validation works by estimating the maximum radius for the planet candidate when accounting for third-light contamination from possible unresolved stars.
If this upper radius limit is below the theoretical radius limit of a brown dwarf ($\sim 0.8\,\rjup$, \citealt{Burrows2011}), the candidate can be securely treated as a planet.

Without contamination, a planet candidate's radius, $R_\mathrm{p}$, is directly related to the planet-star radius ratio, $k$, and stellar radius, $R_\star$, as $R_\mathrm{p} =  k R_\star$. The radius ratio is related to the area ratio, $k^2$, and transit depth, $\Delta F$, as $k = \sqrt{k^2} \sim \sqrt{\Delta F}$, and can be estimated with the help of a transit model that also accounts for the effects from the stellar limb darkening and the planet's orbital geometry.

Third-light contamination from unresolved sources inside a photometric aperture dilutes a transit signal, making a transit with a \enquote{true} depth, $\Delta F_\mathrm{true}$, to appear to have an \enquote{apparent} depth of
\begin{equation}
\Delta F_\mathrm{app} = c + (1-c) \Delta F_\mathrm{true},   
\end{equation}
where $c$ is the contamination, $c = F_\mathrm{c} / (F_\mathrm{c} + F_\mathrm{h})$, $F_\mathrm{c}$ is the flux from the contaminants, and $F_\mathrm{h}$ is the flux from the candidate host.
The diluted transit depth results in an underestimated radius ratio, and, consequently, an underestimated planet candidate's radius. In an extreme case, a strongly contaminated eclipsing binary can appear as a small planet when observed in a single passband.

Contamination depends on the spectral type of the contaminating stars, observation passband, instrument pixel size and point spread function (PSF), and photometry aperture. Consequently, the apparent transit depth also varies between instruments and passbands.

The passband dependency allows for estimating contamination within an aperture centred around the host star using multicolour transit observations. Our multicolour contamination analysis integrates a physical contamination model with a transit model. The physical contamination model, parameterised by the effective temperatures of the planet candidate host star and the contaminating stars, and the contamination factor in some reference passband, calculates the passband-integrated contamination factors based on theoretical stellar spectra by \citet{Husser2013}. These factors are then used to dilute the transits created by the transit model. Marginalising over all the host and contaminant star temperatures and reference contamination levels allowed by the photometry gives us a robust estimate for the planet-star radius ratio.

Unlike in our previous papers \citep{Parviainen2019, Parviainen2020, Parviainen2021, Esparza-Borges2022, Morello2023}, in this paper, we distinguish the \enquote{robust} radius and area ratio estimates from the \enquote{true} ratios. The \enquote{robust} ratios refer to the estimates inferred from the observations when using a model that accounts for the contamination; the \enquote{apparent} ratios refer to the estimates that would be inferred with a model that does not account for possible contamination; and the \enquote{true} ratios refer to the actual, unknown, true geometric ratios. The apparent ratio posteriors contain the true ones if no contamination is present (and the systematics are modelled correctly) but will be biased in the presence of contamination. The robust ratio estimates have significantly larger uncertainties than the apparent ones, but their posteriors will contain the true ratios in the presence of contamination.

The apparent radius ratio in passband $i$ is related to the true are ratio as
\begin{equation}
    k_\mathrm{i,app} = \ktrue \sqrt{1-c_\mathrm{i,true}},
\end{equation}
where $c_\mathrm{i,true}$ is the actual contamination in the passband, and the robust radius ratio is related to the apparent one as
\begin{equation}
    k_\mathrm{rob} =  k_\mathrm{i,app} / \sqrt{1-c_\mathrm{i,est}},
\end{equation}
where $c_\mathrm{i,est}$ is the contamination estimate in the passband. The robust and true radius ratios do not depend on the passband or the instrument, but the apparent radius ratios and contamination factors do.

While the passband-specific variation can be explained using a physical model, the instrumental variation cannot. However, this is not a major issue because the ground-based observations generally have a similar spatial resolution. The only major difference is between the ground-based instruments and \tess because the \tess photometry has significantly lower spatial resolution than the ground-based photometry due to \tess's large pixel size. This means that the \tess photometry cannot generally be modelled using the same physical model as the ground-based photometry. Instead, we include it in the analysis parameterised by an independent apparent area ratio parameter. 

The final multicolour photometry dataset consists of the 55 transit light curves observed with \tess, \hipercam, MuSCAT2, MuSCAT3, and LCOGT 1\,m telescopes. We calculate priors on the limb darkening coefficients using \ldtk \citep{Parviainen2015b}. Further, we have assumed zero eccentricity in all the analyses given the short circularisation time scales for short-period planets \citep{Dawson2018}.
	
\begin{table}
\centering
\caption{Transit light curve model parameters and priors.}
\label{table:model_parameters}
\begin{tabular*}{\columnwidth}{@{\extracolsep{\fill}} llll}
\toprule\toprule
Description & Parameter & Units & Prior \\
\midrule     
\multicolumn{4}{l}{\emph{Global parameters}} \\
\midrule
Zero epoch & $T_0$ & [BJD] & N$^a$ \\
Orbital period & $P$ & [d] & N$^a$ \\
Stellar density  & $\rho$ & [\gcm] & U(5, 35) \\
Impact parameter & $b$    &  & U(0, 1) \\
Apparent area ratio & $k^2_\mathrm{app}$  &  & U(0.02$^2$, 0.08$^2$) \\
Apparent area ratio & $k^2_\mathrm{app,TESS}$  &  & U(0.02$^2$, 0.08$^2$)$^b$ \\
Robust area ratio & $k^2_\mathrm{true}$  &  & U(0.02$^2$, 0.95$^2$) \\
Host temperature & $T_\mathrm{eff,h}$  & [K] & N(3200, 160) \\
Cont. temperature & $T_\mathrm{eff,c}$  & [K] & U(2500, 12000) \\
\\
\multicolumn{4}{l}{\emph{Passband-dependent parameters}$^c$} \\
\midrule
Power-2 $h_1$ in \tess & $h_{1,\tess}$ &  & N(0.78, 0.008) \\
Power-2 $h_2$ in \tess & $h_{2,\tess}$ &  & N(0.69, 0.124) \\
Power-2 $h_1$ in $g$   & $h_{1,g}$ &  & N(0.64, 0.014) \\
Power-2 $h_2$ in $g$   & $h_{2,g}$ &  & N(0.61, 0.070) \\
Power-2 $h_1$ in $r$   & $h_{1,r}$ &  & N(0.65, 0.015) \\
Power-2 $h_2$ in $r$   & $h_{2,r}$ &  & N(0.56, 0.079) \\
Power-2 $h_1$ in $i$   & $h_{1,i}$ &  & N(0.74, 0.012) \\
Power-2 $h_2$ in $i$   & $h_{2,i}$ &  & N(0.68, 0.131) \\
Power-2 $h_1$ in $z_s$ & $h_{1,z_s}$ &  & N(0.79, 0.011) \\
Power-2 $h_2$ in $z_s$ & $h_{2,z_s}$ &  & N(0.71, 0.155) \\
\\
\multicolumn{4}{l}{\emph{Light-curve-dependent parameters}} \\
\midrule
Log$_{10}$ white noise & $\log_{10} \sigma$ &  & U(-4, 0) \\
Baseline coefficient & $s_i$ &  & N$^e$ \\
\bottomrule
\end{tabular*}
\tablefoot{
    The global parameters are independent of the passband or light curve, the passband-dependent parameters are repeated for each passband, and the light-curve-dependent parameters are repeated for each separate light curve. N$(\mu, \sigma)$ stands for a normal prior with a mean $\mu$ and standard deviation $\sigma$, U$(a,b)$ stands for a uniform distribution from $a$ to $b$, and \enquote{Cont. temperature} stands for the effective temperature of the contaminating star.\\
    \tablefoottext{a}{The zero epoch is given a normal prior N(2459255.694, 0.015) and the period is given a normal prior N(2.3262, 0.0002).}
    \tablefoottext{b}{The \tess transits are given a separate apparent radius ratio that effectively makes the \tess contamination independent of the contamination in the ground-based observations.}
    \tablefoottext{c}{The limb darkening coefficients correspond to the transformed power-2 limb darkening law coefficients \citep{Maxted2018} and have normal priors calculated using \ldtk.}
    \tablefoottext{d}{The average log$_{10}$ white noise parameters for each light curve have uninformative uniform priors.}
    \tablefoottext{e}{The linear baseline model coefficients have loose normal priors based on the light curve variability. We do not write them here explicitly, but they can be found from the \texttt{02\_joint\_analysis.ipynb} notebook in the project's GitHub repository.}
    }
\end{table}

We deviate from the analyses in \citet{Parviainen2019}, \citet{Parviainen2020}, and \citet{Parviainen2021} by using a transit model following the power-2 limb darkening law \citep{Morello2017, Maxted2018, Maxted2019} implemented by the \texttt{RoadRunner} transit model in \pytransit \citep{Parviainen2015,Parviainen2020b}. Otherwise, the analysis follows the steps described in these previous papers. The posterior estimation begins with a global optimisation run using the Differential Evolution global optimisation method \citep{Storn1997, Price2005} that results in a population of parameter vectors clumped close to the global posterior mode. This parameter vector population is then used as a starting population for the MCMC sampling with \textsc{emcee}, and the sampling is carried out until a suitable posterior sample has been obtained \citep{Parviainen2018}.	The model parametrisation, priors, and the construction of the posterior function follow directly \citet{Parviainen2020}, and are listed in Table~\ref{table:model_parameters}.

The analyses were carried out with a custom Python code based on \pytransit~v2\footnote{\url{https://github.com/hpparvi/pytransit}} \citep{Parviainen2015, Parviainen2019, Parviainen2020b}, which includes a physics-based contamination model based on the \textsc{PHOENIX}-calculated stellar spectrum library by \citet{Husser2013}. The limb darkening computations were carried out with \ldtk$\!$\footnote{\url{https://github.com/hpparvi/ldtk}} \citep{Parviainen2015b}, and Markov Chain Monte Carlo	(MCMC) sampling was carried out with \textsc{emcee} \citep{Foreman-Mackey2012, Goodman2010}. The code relies on the	existing \textsc{Python} packages for scientific computing and astrophysics: \textsc{SciPy}, \textsc{NumPy} \citep{VanderWalt2011}, \textsc{AstroPy} \citep{TheAstropyCollaboration2013,Price-Whelan2018}, \textsc{photutils} \citep{Bradley2022}, \textsc{astrometry.net} \citep{Lang2010}, \textsc{IPython} \citep{Perez2007}, \textsc{Pandas} 	\citep{Mckinney2010}, \textsc{xarray} \citep{Hoyer2017}, \textsc{matplotlib} \citep{Hunter2007}, and \textsc{seaborn}. The code and the data are publicly available from	GitHub\footnote{\url{https://github.com/hpparvi/parviainen_2021_toi_2266}} as Jupyter notebooks.

\subsection{Results from multicolour validation and system characterisation}
\label{sec:transits.results}

\begin{figure*}
    \centering
    \includegraphics[width=\textwidth]{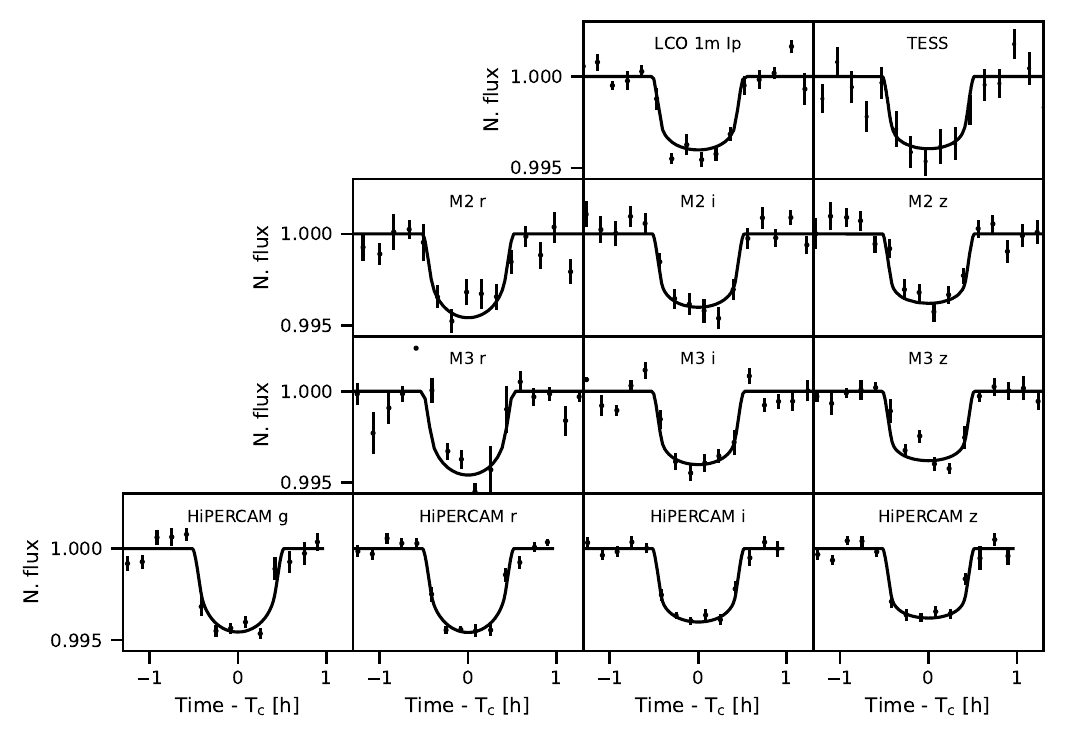}
    \caption{\tess, LCO 1m, MuSCAT2, MuSCAT3, and \hipercam light curves together with the posterior median models. The median posterior baseline model has been removed from the observed photometry, and the observations have been combined, phase folded and binned to 10 minutes for each instrument and passband for visualisation.}
    \label{fig:light_curves}
\end{figure*}

\begin{figure*}
    \centering
    \includegraphics[width=\textwidth]{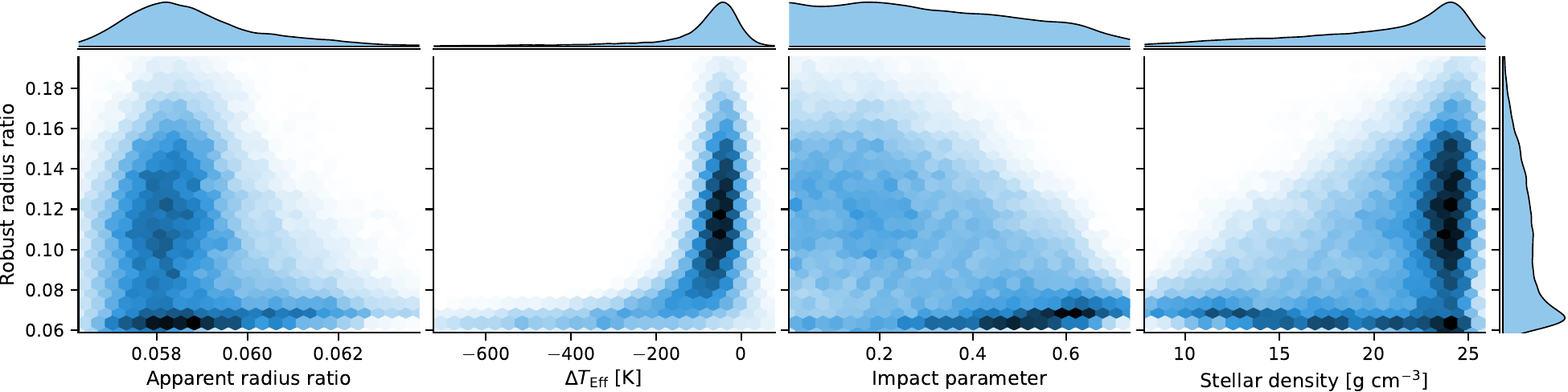}
    \caption{Marginal and joint posterior distributions for the robust radius ratio, apparent radius ratio, difference between the host and contaminant effective temperatures, impact parameter, and stellar density from the multicolour contamination analysis.}
    \label{fig:contamination_posteriors}
\end{figure*}

\begin{table*}
    \centering
    \small
    \caption{Relative and absolute estimates for the stellar and companion parameters derived from the
             multicolour transit analysis.}
    \begin{tabular*}{\textwidth}{@{\extracolsep{\fill}} llll}        
        \toprule\toprule
        Description & Parameter & Units & Posterior \\
        \midrule     
        \multicolumn{4}{l}{\emph{Ephemeris}} \\
        \midrule
        Zero epoch & $T_0$ & [BJD] & $ 2459255.6948195 \pm 2.7 \times 10^{-4}$\\
        Orbital period & $P$ & [days] &  $2.3263180 \pm  4.8\times 10^{-6}$ \\
        Transit duration & $T_{14}$ & [h] & $0.98 \pm 0.01$ \\
        \\
        \multicolumn{4}{l}{\emph{Contamination-analysis related properties}} \\
        \midrule
        Apparent area ratio in the ground-based data & $k^2_\mathrm{app}$  & & $0.0034 \pm 0.0002$ \\
        Apparent area ratio in \tess data & $k^2_\mathrm{app,TESS}$  &  & $0.0034 \pm 0.0004$ \\
        Robust area ratio & $k^2_\mathrm{true}$  &  & < 0.035 (99th percentile upper limit) \\
        Host temperature & $T_\mathrm{eff,h}$  & [K] & $3200 \pm 170$ \\
        Contaminant temperature & $T_\mathrm{eff,c}$  & [K] & $3100 \pm 220$ \\ 
        \\
        \multicolumn{4}{l}{\emph{Relative properties}} \\
        \midrule
        Radius ratio &$k_\mathrm{app}$ & $[R_\star]$ & $ 0.058 \pm 0.001$ \\
        Scaled semi-major axis &$a_\mathrm{s}$ & $[R_\star]$ & $18.43\;(-1.5)\;(+0.7)$  \\
        Impact parameter &$b$ && $ < 0.64$ \\
        \\
        \multicolumn{4}{l}{\emph{Absolute properties}} \\
        \midrule 
        Radius$^a$ &$R_{\mathrm{p,app}}$ & [\rearth]  &  $ \rmedian \pm \rerror $ \\
        Semi-major axis$^a$& $a$ &[AU] & $ 0.020 \pm 0.002 $\\
        Eq. temperature$^b$ & $T_{\mathrm{eq}}$ &[K] & $ 550 \pm 47 $ \\
        Stellar density & $\rho_\star$ & $[\gcm]$ & $22 \; (-4.9) \; (+2.5) $\\
        Inclination & $i$ &[deg] & $> 88.51$ \\
        Bolometric insolation & S & [S$_\oplus$] & $13\pm3.0$\\
        \bottomrule       
    \end{tabular*}
    \tablefoot{ The estimates
        correspond to the posterior median ($P_{50}$) with $1 \sigma$ uncertainty estimate based on the 16th and 84th posterior percentiles ($P_{16}$ and $P_{84}$, respectively) for symmetric, approximately normal posteriors. For asymmetric, unimodal posteriors, the estimates are $P_{50}{}^{P_{84}-P_{50}}_{P_{16}-P_{50}}$. 
        \tablefoottext{a}{The semi-major axis and planet candidate radius are based on the scaled            semi-major axis and apparent radius ratio samples, and the stellar radius estimate shown in Table~\ref{tbl:star}.} 
        \tablefoottext{b}{The equilibrium temperature of the planet candidate is calculated using the stellar \teff estimate, scaled semi-major axis distribution, heat redistribution factor distributed uniformly between 0.25 and 0.5, and planet's Bond albedo distributed uniformly between 0 and 0.4.}}
    \label{table:parameters}  
\end{table*}

We show the photometry used in the multicolour analysis with the posterior transit model in Fig.~\ref{fig:light_curves}, and the posterior densities for the true radius ratio, the effective temperature of the contaminant, impact parameter, and stellar density in Fig.~\ref{fig:contamination_posteriors}. The multicolour analysis robustly rejects any false positive scenarios where the transit signal would not be caused by a small transiting planet with a false alarm probability (FAP) equivalent to $0$ (see Sect.~\ref{sec:discussion.summary} for more details about the multicolour validation results). 

Since false positive scenarios affecting the planet candidate radius significantly can be rejected, we adopt the values from a separate uncontaminated light curve analysis as our final system characterisation results. The posterior estimates for the stellar and planetary parameters inferred from the analysis ignoring possible contamination are listed in Table~\ref{table:parameters}.   

\subsection{Transit timing variations}
\label{sec:transits.ttvs}

We carried out an additional transit timing variation analysis where the transit centre times for each transit were free parameters in the model. The transit centres from the \tess observations were poorly constrained due to the low SNR for a single transit. Still, the ground-based observations reached up to 30-second 1$\sigma$ transit centre precision and agreed with a linear period without signs of significant dynamical interactions with other possible bodies in the system.

\section{Discussion}
\label{sec:discussion}

\subsection{Validation summary}
\label{sec:discussion.summary}

We validate \tplanet as a small planet with a radius of $\rmedian\pm\rerror~\rearth$ based on high-resolution imaging and multicolour transit photometry. Specifically,
\begin{itemize}
    \item high-resolution imaging rules out significant blending from sources with angular separation $\gtrsim 0.25\arcsec$,
    \item and multicolour transit photometry rules out significant contamination from stars of different spectral type than the host.
\end{itemize}

The analysis rules out any contamination from sources dissimilar to the host star that would significantly affect the radius of the transiting object (see the $\Delta T_\mathrm{Eff}$ vs robust radius ratio posterior in Fig.~\ref{fig:contamination_posteriors}). This applies equally to scenarios where the transit signal would occur in a faint background star (in which case our spectroscopic characterisation of the host star would be incorrect) or where the signal would occur in the assumed host star but would be blended with fainter contaminant stars.

For $\Delta T_\mathrm{Eff} \approx 0$, the true radius ratio and contamination are constrained by the achromatic transit geometry: radius ratios larger than the inferred upper limit cannot produce the observed flat-bottomed transit with relatively short ingress and egress durations. The multicolour analysis yields a true radius ratio 99th percentile posterior upper limit of 0.19. Since the host star needs to be similar to the assumed host star, the upper radius ratio limit corresponds to a contamination factor of 90\% and a planetary radius of 5\rearth that is well below the brown dwarf radius limit of 0.8\rjup. However, a contamination level this high is physically implausible because the scenario would require $\sim$10 M~dwarfs with $\teff\sim3200$~K (one being orbited by the transiting object) residing within angular separations $\lesssim 0.5\arcsec$ of each other (see Fig.~\ref{fig:high_res_imaging}). 

The only realistic contamination scenario not rejected by the analysis would be that the transiting object orbits a component of an equal-mass binary. This would lead to a contamination factor of 50\% and a planet radius of 2\rearth.

\subsection{Radius uncertainty}
\label{sec:discussion.radius_uncertainty}

Our radius ratio estimate for \tplanet has a relative uncertainty of 2\%, while the absolute radius estimate has a relative uncertainty of 4.3\%. The high precision in radius ratio is largely thanks to the transit observed with \hipercam, while the significantly lower precision in the absolute radius is due to the 3.8\% relative uncertainty in the stellar radius. M~dwarf radii and masses are notoriously challenging to estimate reliably, and the uncertainty in the stellar radius impedes any attempts to improve the absolute planet radius estimate by observing additional transits.

\subsection{\tplanet and the M~dwarf radius valley}
\label{sec:discussion.radius_valley}

\begin{figure}[t]
    \centering
    \includegraphics[width=\columnwidth]{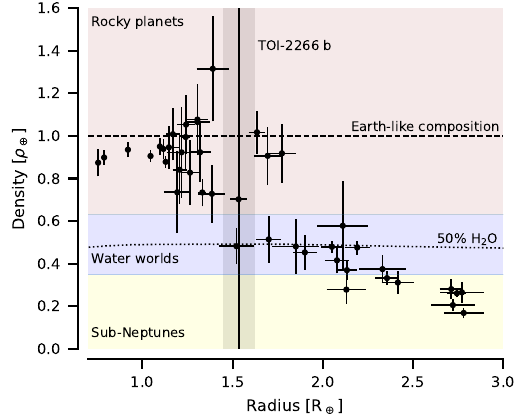}
    \caption{\tplanet's location in the plane of radius and relative density (black vertical line and shading) with well-characterised small planets orbiting M~dwarfs with periods shorter than 32~d by \citet{Luque2022}. The colour shading corresponds to rocky planets (brown), water worlds (blue), and sub-Neptunes (yellow).}
    \label{fig:radius_vs_density}
\end{figure}

\begin{figure*}[t]
    \centering
    \includegraphics[width=\textwidth]{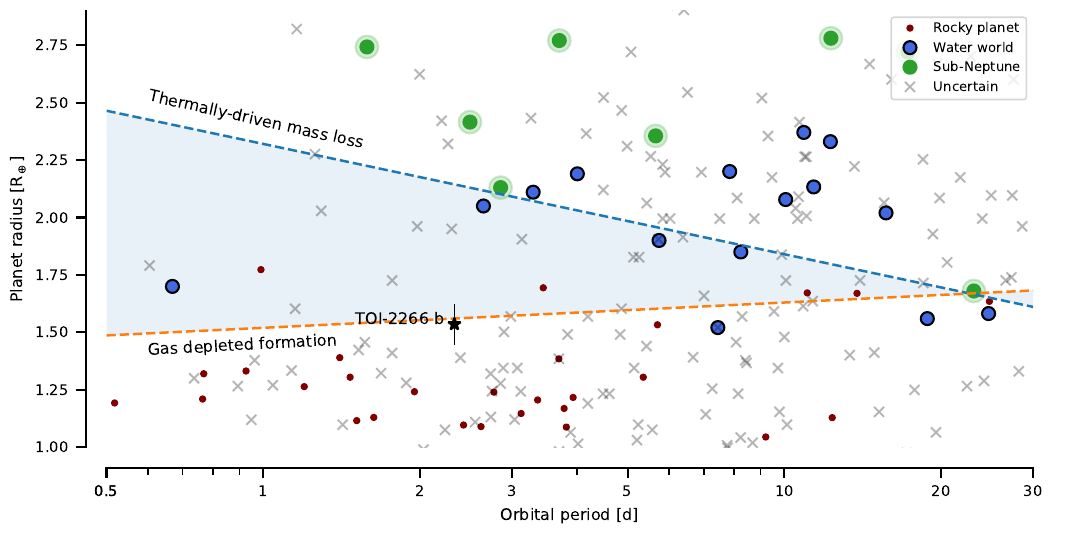}
    \caption{\tplanet's location in the period-radius plane with the currently known planets orbiting M~dwarfs. The planets without sufficiently precise density estimates are shown as grey crosses (From \url{exoplanet.eu}, accessed 2023~April~24), while the planets with well-constrained densities from the catalogue by \citet{Luque2022} are separated by their likely type: rocky planets are shown as black dots, water worlds as blue circles, and sub-Neptunes as green circles. The upper radius limits for rocky planets for the gas-depleted formation and thermally-driven mass loss scenarios are drawn as dashed lines.}
    \label{fig:period_vs_radius}
\end{figure*}

\tplanet's period and radius make it a welcome addition to a relatively small sample of known planets that can be used to study planet formation around M~dwarfs and, especially, probe the transition zone between rocky and water-rich planets. 

First, considering the recent work by \citet{Luque2022}, with a radius of $\rmedian\pm\rerror~\rearth$, \tplanet falls inside a transition zone where we find both rocky and water-rich planets, as shown in Fig.~\ref{fig:radius_vs_density}. This transition zone corresponds to the M~dwarf radius valley in planet radius space \citep{Cloutier2020b, VanEylen2021}, but its extent is poorly defined due to the small number of known planets inside it. If we take the zone to span from 1.5 to 1.8~\rearth (roughly corresponding to the radius of the smallest known water world and the largest known earth-like planet) and consider the planets with masses and radii estimated to a precision of 25\% and 8\% or better,\!\footnote{Based on an updated catalogue by R.~Luque, private communication.} the zone encompasses three rocky planets, four water worlds, and one planet with an intermediate composition. This is insufficient to constrain the extent of the zone or study its physics, but more planets with accurately measured masses are needed. Mass measurement of \tplanet should be within reach of the current instruments (see discussion below), and even an upper mass limit measured to an RV semi-amplitude precision of 4~m/s suffice to determine whether the planet is rocky or water-rich.

Second, considering the M~dwarf radius valley \citep{Cloutier2020b} and rocky planet formation, \tplanet is located in a sparsely populated region in the period-radius plane (Fig.~\ref{fig:period_vs_radius}) inside the \enquote{keystone planet} wedge where rocky planet formation scenarios make disagreeing predictions on how the radius valley location depends on the planet's orbital period \citep{Cherubim2022, Cloutier2021, VanEylen2021, Cloutier2020b}. Simplistically, the thermally-driven mass loss scenario proposes that rocky planets are stripped cores of planets that accreted a significant atmosphere during their birth but lost it due to core-powered mass loss, photoevaporation, or one of several other physical processes that can strip a planet of its atmosphere, while the gas depleted formation scenario proposes that the rocky planets formed later than the planets with significant H/He envelopes after the gas in the protoplanetary disk had dissipated. The thermally-driven mass loss scenario predicts that the upper limit of rocky-planet radii decreases with the orbital period because the process can strip a larger planet of its atmosphere the closer to the star the planet migrates. The gas-depleted scenario predicts an opposite trend where the upper limit for the rocky-planet radius increases slightly with the orbital period because the forming planets can accrete more mass the longer their period is. 

\tplanet is located at the lower end of the \enquote{keystone planet} wedge in the period-radius plane, so estimating its density via RV mass measurements can help shed light on the dominant small-planet formation pathway for M~dwarfs. Were \tplanet to be water-rich (or a sub-Neptune), it would further contribute to the growing evidence that small planets are mainly formed through the gas-depleted formation scenario \citep{Cherubim2022}.

\subsection{Possible photometric signal related to stellar rotation}
\label{sec:discussion.stellar_rotation}

A Lomb-Scargle analysis \citep[LS;][]{Lomb1976, Scargle1982} of the \tess photometry divided by the best-fitting transit model using the Generalized Lomb Scargle periodogram by \citep[GLS,][]{Zechmeister2009} shows evidence for a periodic variability with a period of 4.54~d and a semi-amplitude of $1300 \pm 100$~ppm. The signal is loosely sinusoidal in shape and its period is close to being twice the planet's orbital period ($2P = 4.65$~d). It is unlikely that the signal would be caused by the planet, and we consider it more likely that it is indicative of a stellar rotation period of $\approx 4.5$~d. This would agree well with the results by \citet{Popinchalk2021}, who measured relatively rapid stellar rotation periods of 0.3-10~d for M5 dwarfs of all ages.

\subsection{Prospects for RV follow-up}

\begin{figure}
    \centering
    \includegraphics[width=\columnwidth]{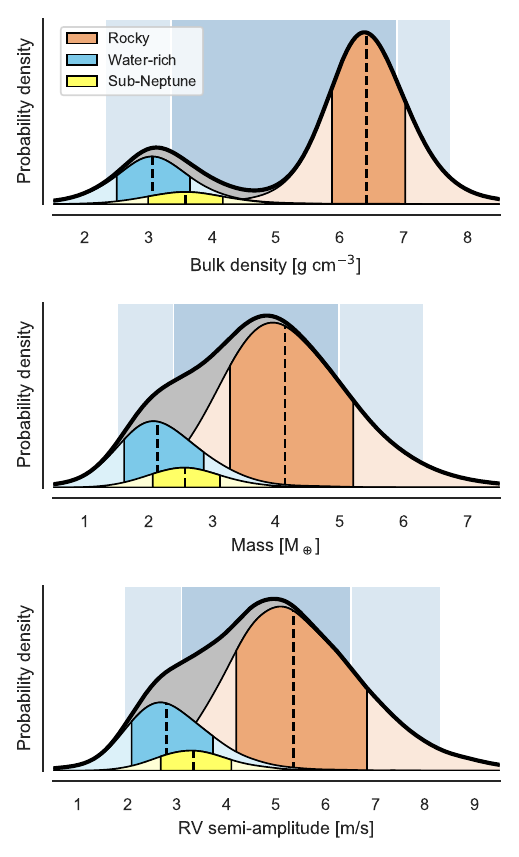}
    \caption{\tplanet's bulk density, mass, and radial velocity semi-amplitude probability distributions given a posterior radius estimate of $\rmedian \pm \rerror\,\rearth$ predicted by the \spright package. The complete probability distribution is marked by a thick black line and gray shading, and the individual contributions from the three \spright model components (rocky planets, water-rich planets, and sub-Neptunes) are plotted in light brown, light blue, and yellow. The blue shading in the background shows the 68\% and 95\% central posterior intervals for the distributions.}
    \label{fig:rv_semi_amplitude}
\end{figure}

\begin{figure}
    \centering
    \includegraphics[width=\columnwidth]{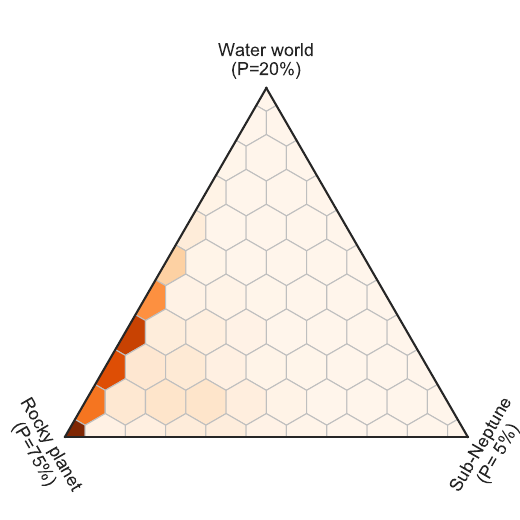}
    \caption{\tplanet's composition class based on its radius predicted by the \spright package.}
    \label{fig:spright_class}
\end{figure}

We use a numerical radius-mass relation provided by the \spright package\footnote{The analysis used \spright version 23.11.01 (\href{https://doi.org/10.5281/zenodo.10082653}{10.5281/zenodo.10082653}); the \spright package is available from \url{https://github.com/hpparvi/spright} and \texttt{PyPI}.} \citep{Parviainen2023a} to predict \tplanet's mass and RV semi-amplitude distributions (Fig.~\ref{fig:rv_semi_amplitude}) and composition class given the planet's radius (Fig.~\ref{fig:spright_class}). 
\spright is a novel probabilistic mass-density-radius relation for small planets that represents the joint planetary radius and bulk density probability distribution as a mean posterior predictive distribution of an analytical three-component mixture model. The three components represent rocky planets, water-rich planets, and sub-Neptunes, and the final numerical probability model is obtained by marginalising over all analytical model solutions allowed by observations. The approach allows for solutions where the water-rich planet component does not exist, and so the final \spright mass-radius model is agnostic to the existence of water-rich planets as a separate population on their own. 

For \tplanet, \spright predicts an RV semi-amplitude, $K$, of $1.9-8.3$~\mps (95\% central posterior limits, Fig.~\ref{fig:rv_semi_amplitude}). If the planet is rocky, we expect a $K$ value of $5.4 \pm 1.3$~\mps, while for water-rich planets and sub-Neptunes we expect $K$ values of $2.8 \pm 0.8$ and $3.3 \pm 0.7$~\mps, respectively.

The RV semi-amplitudes are large enough that the planet's mass can be expected to be measurable using RV observations with the currently available red-sensitive instruments. Due to \tstar's high declination, it makes a poor target for telescopes in the southern hemisphere. However, the star is observable from telescopes located in Mauna Kea, Hawaii, during the summer period, reaching a minimum airmass of 1.02 in late May. Thus, \tplanet would be amenable to mass measurements using \maroonx \citep{Seifahrt2018}, \kpf \citep{Gibson2016}, or SPIRou \citep{Donati2020}.

Assuming good observing conditions and exposure times of one hour, the instrument-specific exposure time calculators\!\footnote{The exposure time calculator for \kpf can be found from \url{https://github.com/California-Planet-Search/KPF-etc}; for \maroonx from \url{http://www.maroonx.science/} and \url{https://www.gemini.edu/instrumentation/maroon-x/exposure-time-estimation}; and for \spirou from \url{https://etc.cfht.hawaii.edu/spi}.} predict RV observation uncertainties of 1~\mps for \maroonx, 2~\mps for \kpf, and 6-11~\mps for \spirou. 
We carried out numerical RV mass measurement simulations to study the precision and significance of a mass measurement achieved by 4, 6, 8, and 10 one-hour exposures with these three instruments considering five composition scenarios corresponding to the 2.5\% and 97.5\% \spright posterior percentiles, and the $K$ posterior median values for the rocky, water-rich and (puffy) sub-Neptune compositions. For a single simulation, we created a set of $N_\mathrm{obs}$ simulated RV observations with observation phases clustered randomly close to the RV signal minima and maxima, $K$ following from a given composition scenario, and noise following the instrument-specific noise estimate for a one-hour exposure. After this, we estimated the posterior distribution for $K$ given the simulated measurements using \pytransit's \texttt{RVLPF} class. We repeated the simulation 10 times for each combination of the composition classes, instruments, and number of exposures, and summarise the average mass measurement significances and precisions\!\footnote{We define the \enquote{mass measurement significance} here as the $K$ posterior median divided by the posterior's standard deviation, that is, the distance of the posterior median from zero in units of standard deviation.} in Table~\ref{table:rv_simulation}. 

The results in Table~\ref{table:rv_simulation} are optimistic because they consider only photon noise and ignore correlated noise from stellar granulation and variability, but they nevertheless allow us to conclude that \tplanet's mass can likely be estimated with a relatively small number of \maroonx or \kpf observations. \maroonx observations of bright M1.5V star GJ~806 \citep{Palle2023} and M3.5~V star Gl~486 \citep{Caballero2022} have lead to additional RV jitter estimates for the \maroonx red arm RV observations up to 1~\mps, while stellar variability and star spots can lead to quasi-periodic RV signals with amplitudes up to 10-20~\mps \citep{Kossakowski2022, Cortes-Zuleta2023}. The short-time-scale jitter should not form a major obstacle for \maroonx observations since additional noise of 1~\mps leads to \kpf-like performance. However, the larger-amplitude RV signals related to stellar variability have periods matching the stellar rotation period, and if the periodic photometric signal of 4.54~d identified in Sect.~\ref{sec:discussion.stellar_rotation} corresponds to the stellar rotation period, a significantly larger number of observations may be required to ensure that the RV signal caused by the planet can be disentangled from the one cause by stellar variability.

\begin{table}
    \centering
    \caption{Simulated ideal RV mass measurement significances and precisions for \maroonx, \kpf, and \spirou.}
    \label{table:rv_simulation}
\begin{tabular*}{\columnwidth}{@{\extracolsep{\fill}} ll|rrrrr}
\toprule
\toprule
Instrument & $N_\mathrm{obs}$ & low & high & rocky & water & puffy \\
\midrule
\\
 \multicolumn{7}{l}{Simulated RV mass measurement significance [$\sigma$]}\\
\midrule
\multirow[c]{4}{*}{MAROON-X} & 4 & 4 & 16 & 11 & 5 & 7 \\
 & 6 & 5 & 20 & 13 & 7 & 8 \\
 & 8 & 5 & 24 & 16 & 7 & 9 \\
 & 10 & 6 & 25 & 17 & 10 & 10 \\
 \midrule
\multirow[c]{4}{*}{KPF} & 4 & 2 & 9 & 5 & 3 & 3 \\
 & 6 & 2 & 10 & 6 & 4 & 4 \\
 & 8 & 3 & 12 & 8 & 4 & 4 \\
 & 10 & 3 & 13 & 8 & 4 & 5 \\
 \midrule
\multirow[c]{4}{*}{SPIRou-low} & 4 & 1 & 3 & 2 & 2 & 2 \\
 & 6 & 2 & 4 & 3 & 2 & 2 \\
 & 8 & 1 & 5 & 3 & 2 & 2 \\
 & 10 & 2 & 5 & 4 & 2 & 2 \\
 \midrule
\multirow[c]{4}{*}{SPIRou-high} & 4 & 2 & 3 & 2 & 2 & 2 \\
 & 6 & 1 & 3 & 2 & 2 & 2 \\
 & 8 & 1 & 3 & 2 & 2 & 2 \\
 & 10 & 2 & 3 & 2 & 1 & 2 \\
 \\
 \multicolumn{7}{l}{Simulated RV mass measurement precision [\%]}\\
\midrule
\multirow[c]{4}{*}{MAROON-X} & 4 & 28 & 6 & 10 & 19 & 15 \\
& 6 & 22 & 5 & 8 & 15 & 12 \\
& 8 & 19 & 4 & 6 & 14 & 11 \\
& 10 & 18 & 4 & 6 & 11 & 10 \\
\midrule
\multirow[c]{4}{*}{KPF} & 4 & 50 & 12 & 20 & 39 & 37 \\
& 6 & 45 & 10 & 16 & 30 & 30 \\
& 8 & 37 & 9 & 13 & 27 & 23 \\
& 10 & 34 & 8 & 12 & 27 & 20 \\
\midrule
\multirow[c]{4}{*}{SPIRou-low} & 4 & 76 & 37 & 61 & 69 & 64 \\
& 6 & 68 & 29 & 49 & 63 & 50 \\
& 8 & 83 & 22 & 41 & 52 & 57 \\
& 10 & 62 & 24 & 32 & 57 & 57 \\
\midrule
\multirow[c]{4}{*}{SPIRou-high} & 4 & 71 & 41 & 60 & 69 & 69 \\
& 6 & 83 & 43 & 56 & 57 & 67 \\
& 8 & 80 & 40 & 52 & 72 & 55 \\
& 10 & 63 & 44 & 57 & 79 & 71 \\
\bottomrule
\end{tabular*}
\tablefoot{The significance of an RV mass measurement is here defined as $\mathrm{med}(k)/\sigma(k)$, where med($k$) is the median of the RV semi-amplitude posterior estimate and $\sigma(k)$ its standard deviation, while the RV measurement precision is $\sigma(k)/\mathrm{med}(k)$. The number of 1~h long exposures used in the simulations is marked by $N_\mathrm{obs}$, and low, high, rocky, water, and puffy refer to different planet mass scenarios leading to RV semi-amplitudes of 1.9, 8.3, 5.4, 2.8, and 3.3~\mps, respectively. For \spirou, we consider the lower and upper precision limits of 6 and 11~\mps per exposure separately. The estimates should be considered somewhat optimistic since they consider only photon noise and do not include the RV signals due to stellar granulation and variability.}
\end{table}

\section{Conclusions}

We have validated \tplanet as a small planet (most likely a super-Earth or a water world) using multicolour transit photometry and high-resolution imaging. The planet is amenable to ground-based RV mass estimation with \maroonx and \kpf, and a mass measurement combined with our radius estimate precision of 4\% will make the planet a valuable addition in studying small-planet populations and planet formation scenarios. 

Considering the planet's radius, \tplanet is a welcome addition to a small population of planets located inside a transition zone where \citet{Luque2022} find both rocky planets and water worlds, and measuring the planet's density may allow us to understand better the differences in the formation histories of these two populations. Further, considering rocky planet formation scenarios, the planet occupies a currently sparsely populated region in the period-radius plane, the so-called \enquote{keystone} wedge as defined by \citet{Cloutier2020b}. Were \tplanet to be identified as a water world or a sub-Neptune, this would increase support for the gas-depleted formation scenario.

\begin{acknowledgements}
We thank the anonymous referee for their extremely helpful comments and suggestions.
We acknowledge financial support from the Agencia Estatal de Investigación of the Ministerio de Ciencia, Innovación y Universidades and the European FEDER/ERF funds through projects ESP2013-48391-C4-2-R, AYA2016-79425-C3-2-P, AYA2015-69350-C3-2-P, and PID2019-109522GB-C53, and PGC2018-098153-B-C31.  
This work is partly financed by the Spanish Ministry of Economics and Competitiveness through project ESP2016-80435-C2-2-R.
Funding from the University of La Laguna and the Spanish Ministry of Universities is acknowledged.
HP acknowledges support by the Spanish Ministry of Science and Innovation with the Ramon y Cajal fellowship number RYC2021-031798-I.
This work is partly supported by JSPS KAKENHI Grant Numbers JP15H02063, JP17H04574, JP18H05442, JP18H05439, JP20K14521, JP21K13975, JP21K13955, JP21K20376, JP22000005, JST CREST Grant Number JPMJCR1761, and Astrobiology Center SATELLITE Research project AB022006.
R.L. acknowledges funding from the University of La Laguna through the Margarita Salas Fellowship from the Spanish Ministry of Universities ref. UNI/551/2021-May 26, and under the EU Next Generation funds. 
J.~K. gratefully acknowledges the support of the Swedish National Space Agency (SNSA; DNR 2020-00104) and of the Swedish Research Council  (VR: Etableringsbidrag 2017-04945).
G.N. thanks for the research funding from the Ministry of Education and Science programme the \enquote{Excellence Initiative - Research University} conducted at the Centre of Excellence in Astrophysics and Astrochemistry of the Nicolaus Copernicus University in Toru\'n, Poland. 
The postdoctoral fellowship of KB is funded by F.R.S.-FNRS grant T.0109.20 and by the Francqui Foundation.
This publication benefits from the support of the French Community of Belgium in the context of the FRIA Doctoral Grant awarded to MT.
This article is partly based on observations made with the MuSCAT2 instrument, developed by ABC, at Telescopio Carlos S\'anchez operated on the island of Tenerife by the IAC in the Spanish Observatorio del Teide.
Based on observations made with the Nordic Optical Telescope, owned in collaboration by the University of Turku and Aarhus University, and operated jointly by Aarhus University, the University of Turku and the University of Oslo, representing Denmark, Finland and Norway, the University of Iceland and Stockholm University at the Observatorio del Roque de los Muchachos, La Palma, Spain, of the Instituto de Astrofisica de Canarias.
The data presented here were obtained in part with ALFOSC, which is provided by the Instituto de Astrofisica de Andalucia (IAA) under a joint agreement with the University of Copenhagen and NOT.
We acknowledge the use of public \tess Alert data from pipelines at the \tess Science Office and at the \tess Science Processing Operations Center. 
This work makes use of observations from the LCOGT network. Part of the	LCOGT telescope time was granted by NOIRLab through the Mid-Scale Innovations Program (MSIP). MSIP is funded by NSF. 
This paper is based on observations made with the MuSCAT3 instrument, developed by the Astrobiology Center and under financial supports by JSPS KAKENHI (JP18H05439) and JST PRESTO (JPMJPR1775), at Faulkes Telescope North on Maui, HI, operated by the Las Cumbres Observatory. 
This paper made use of data collected by the TESS mission and are publicly available from the Mikulski Archive for Space Telescopes (MAST) operated by the Space Telescope Science Institute (STScI). Funding for the TESS mission is provided by NASA’s Science Mission Directorate.
Resources supporting this work were provided by the NASA High-End Computing (HEC) Program through the NASA Advanced Supercomputing (NAS) Division at Ames Research Center for the production of the SPOC data products.
This work made use of \texttt{tpfplotter} by J. Lillo-Box (publicly available in
\url{www.github.com/jlillo/tpfplotter}), which also made use of the python packages \texttt{astropy},
\texttt{lightkurve}, \texttt{matplotlib} and \texttt{numpy}.
This work has made use of data from the European Space Agency (ESA) mission
{\it Gaia} (\url{https://www.cosmos.esa.int/gaia}), processed by the {\it Gaia}
Data Processing and Analysis Consortium (DPAC,
\url{https://www.cosmos.esa.int/web/gaia/dpac/consortium}). Funding for the DPAC
has been provided by national institutions, in particular the institutions
participating in the {\it Gaia} Multilateral Agreement.

\end{acknowledgements}

\bibliographystyle{aa} 
\bibliography{toi_2266}
	
\end{document}